\documentclass[a4paper,12pt,aps]{revtex4-1}

\usepackage{amsmath}
\usepackage{bbm}
\usepackage{graphicx,amssymb,bm,latexsym,color,epsf}
\usepackage{breqn}
\usepackage[colorlinks]{hyperref} 

\newcommand{\be}{\begin{equation}}
\newcommand{\ee}{\end{equation}}
\newcommand{\bea}{\begin{eqnarray}}
\newcommand{\eea}{\end{eqnarray}}

\makeatletter
\let\cat@comma@active\@empty
\makeatother

\begin{document}
\title{Lorentz symmetry violating  Lifshitz-type field theories.}

\author{Emiliano Rizza}
\email{emilianorizza72@gmail.com}
\affiliation{Dipartimento di Fisica, Universit\`a di Catania,
\\via S. Sofia 64, I-95123, Catania, Italy}

\author{Dario Zappal\`a}
\email{dario.zappala@ct.infn.it}
\affiliation{INFN, Sezione di Catania, via S. Sofia 64, I-95123, 
Catania, Italy}

\begin{abstract}
\vskip 30pt
\centerline{ABSTRACT}
\vskip 10pt
We discuss the ultraviolet sector of 3+1 dimensional Lifshitz-type anisotropic higher derivative scalar, 
fermion and gauge field theories, with anisotropy exponent z=3 and with explicit breaking of Lorentz symmetry. 
By discarding from the action all momentum dependent vertex operators, which is essential to 
avoid phenomenologically unacceptable deformations of the light cone, we find that renormalizable scalar 
self-interaction and Yukawa-like  couplings are, in general, asymptotically free.
However, the requirement of cancelling momentum dependent vertex operators is incompatible with 
gauge symmetry and, therefore, for this kind of theories, gauge symmetry as well as Lorentz symmetry
are recovered only as emergent properties  below some energy scale $M$, that must be constrained  
from experiments. The quantum corrections to the scalar mass and their impact on the hierarchy problem
are also analyzed.
\end{abstract} 

\maketitle

\setcounter{page}{2}

\section{Introduction\label{sect1}}	

The smoothening of ultraviolet (UV) divergences with the consequent improvement of the renormalizability of a quantum field  theory,
 induced by an enhanced number of space and time derivatives of the field, is a long studied mechanism, especially in the context of 
field theoretical formulation of gravity. \cite{thirring,Pais:1950za,Stelle:1976gc,deUrries:1998obu,Hawking:2001yt,Collins1}

As might be expected, the improved renormalization can be traced back to the presence of a novel Renormalization Group fixed point
and, in fact, fixed points of higher derivative field theories are known in condensed matter as Lifshitz points. \cite{Horn,hornrev,selke1988,Diehl,Diehl3}
They  can be classified as isotropic, if the number of derivatives is the same for every coordinate of the Euclidean space under investigation,
or as anisotropic points, if  there is an anisotropy  between  two sets of coordinates, consisting in a  different number of derivatives with respect 
to the coordinates of one set and  the number of derivatives of the other set.

When (a Wick-rotated version of) a Minkowskian quantum field is under investigation, a manifestly Lorentz invariant action is typically expected, 
mainly because astrophysical observations pose the energy scale of Lorentz violating effects above $10^{10}$ GeV\cite{ellis:2003,chenhuang} 
(or even above $10^{17}$ GeV,\cite{Ellis:2018lca}  if the specific  theory investigated predicts a correction to the dispersion relation that 
grows linearly with the energy).  

Actually, Lorentz symmetric, isotropic Lifshitz points, possess a rich phase structure and are potentially  
useful for the investigation of some  high energy physics issues. \cite{Bonanno:2014yia,Zappala:2017vjf,Zappala:2018khg,Zapp,Defenu,Buccio} 
However, isotropic higher derivative theories  that contain more than two time derivatives  are affected  by the Ostrogradski instability, 
associated with  violation of unitarity. \cite{deUrries:1998obu,Woodard:2006nt} Therefore, such isotropic higher derivative theories
can only be regarded as low energy effective theories of some more fundamental high energy theory that preserves unitarity,
while the only Lifshitz points entitled to describe the UV completion of a fundamental field theory are those with  anisotropic structure, 
generated by actions with two derivatives with respect to the time coordinate and $2z$ derivatives with respect to the space coordinates
(below, we refer to $z$ as the anisotropy exponent).
Clearly, this kind of actions do not respect Lorentz symmetry and one must be careful that their predictions are not in contradiction with  the 
observational constraints and the symmetry is recovered at low energy as an emergent property.

The original formulation of a renormalizable anisotropic gravitational theory, known as  Ho\v{r}ava-Lifshitz gravity \cite{horava}, 
is constructed on the basis of these motivations (for a recent version of this formulation see \cite{Barv1,Barv2,Barv3,Barv4} ), 
and also  anisotropic field  theories on flat spacetime were formulated on the same grounds. 
\cite{Anselmi:2007ri,Iengo,Dhar:2009dx,horava:ym,eune,alexandre,Kikuchi,Chao:2009tqf,Solomon:2017nlh}

In this Letter, we focus on  $3+1$-dimensional field theories,  to investigate on the possibility of constructing a renormalizable, 
consistent and UV safe completion of scalar, as well as fermion and gauge theories on flat spacetime. 
In particular, we focus on the case with  anisotropy exponent $z=3$, rather than $z=2$, in order to achieve an enhanced suppression 
of the UV loop divergences from the modified propagators. As discussed below, for a $3+1$-dimensional system with $z=3$, 
the scaling dimension of the scalar field vanishes and becomes negative for $z>3$. Therefore, we select $z=3$ to avoid theories 
with negative scaling dimension of the field.

More in detail, we  generalize to non-zero spin fields the approach developed  in \cite{zappalast} for scalar fields.
In fact,  a particular restriction on the scalar action is enforced  in \cite{zappalast} to overcome a 
serious inconvenience,  outlined in \cite{Iengo,alexandre}, generated by a class of operators that produce 
dissimilar corrections to the  effective light cones seen by different particles, when these interact. This unacceptable 
effect is actually avoided by the restriction discussed in \cite{zappalast} and, below, we review its 
application to the case of scalar  fields in Section \ref{sect2}, while  the case of  fermion  and gauge fields  
is discussed respectively in  Sections \ref{sect3} and \ref{sect4}.
Conclusions are reported in Section \ref{sect5}.

\section{Scalar Fields\label{sect2}}

The scaling dimension of a scalar field in a 3+1 dimensional Lifshitz-type theory and generic $z$, is easily read from the derivative sector of the action 
\begin{equation}
S_D=\int\;d^3 x \,dt \left \{ \frac{1}{2} \left (\partial_t \phi \right)^2 -  \sum_{k=1}^{z} \,  \sum_{i=1}^{3}  
\frac{a_k}{2 \,M^{2k-2}}  \left ( \, [\partial_i]^k  \phi \right)^2 \right \}
 \, ,
\label{ssk}
\end{equation}
by requiring that, under scale transformations with scale factor $b>1$, the space-time coordinates transform as $x^i \to b  x^i$ and $t \to b^z t $,
which  defines the following scaling dimensions $[x^i]_s=-1$ and $[t]_s=-z$ and, accordingly, the scaling dimension of the field, $[\phi]_s=(3-z+\eta)/2$,
and that of the coefficient of the highest derivative operator,  $[a_z]_s=-\eta$ 
(in these relations we also took into  account  a non-zero anomalous dimension of the field, $\eta$). 
Suitable powers of a mass scale, $M$ are inserted in  Eq.~(\ref{ssk}) to keep track of the canonical 
dimensions of the various operators and couplings, so that all $a_i$  are dimensionless pure numbers.

Clearly, if $z=3$ and $\eta$ is neglected, one finds $[\phi]_s=0$ and this determines  the structure of the interaction sector, which, added to $S_D$, 
represents the complete renormalizable action:
\begin{equation}
S_I= -\,
\int\;d^3 x \,dt \left \{ \sum_{n=2}^{\infty} \frac{g_n \, \phi^n }{n! \, M^{(n-4)}} 
+ 
\sum_{k=1}^{3} \,  \sum_{i=1}^{3}  
\left [  \sum_{m=1}^{\infty}  \frac{w_{m,k}\, \phi^m }{ M^{2k-2+m}}  \right ] \;
\left ( \, [\partial_i]^k  \phi \right)^2 \right \}
 \, .
\label{ssi}
\end{equation}
Since $[\phi]_s=0$ , the potential term, which  
corresponds to the  first sum in curly brackets in (\ref{ssi}), must  include all powers of $\phi$.
For the same reason, the complete renormalizable action must contain the second block in curly brackets,
which is constructed by multiplying the derivative part sector in $S_D$ by any power of $\phi$,
while any other term containing higher powers of the derivatives is,  by construction, non-renormalizable.
The powers of the scale $M$ are again selected so that all couplings $g_n$ and $w_{m,k}$ have zero
canonical dimension.

As pointed out in \cite{Iengo} and, later, analyzed again in \cite{alexandre}, loop integrals 
containing momentum dependent vertices proportional to $w_{m,k}$, produce corrections to the coupling $a_1$,
which is related to the ratio of  space and time coordinates and therefore defines the value of the speed of light.
However, in the case of two  interacting fields, these corrections to $a_1$ turn out to be different for each field,
and one concludes  that the excitations of the two fields have different light cones. 

To avoid this flaw, instead of resorting to a quite unnatural "{\it ad hoc}" tuning of the various $w_{m,k}$, 
it is suggested to restrict the form of the action and  discard all momentum dependent 
vertices, by taking $w_{m,k}=0$ for any $m$ and $k$, even though renormalizable.\cite{zappalast} 
In fact, not only the suppression of 
all momentum dependent vertices avoids  the dangerous corrections to the speed of light, but  one also finds that
such vertices are not generated by quantum corrections associated  to  the other remaining vertices in the action.
To illustrate this point we analyze the divergences of the quantum corrections produced by the action $S_D+S_I$ with $z=3$ and  all $w_{m,k}=0$.  
\begin{figure}[tph]
\centerline{\includegraphics[width=4.0in]{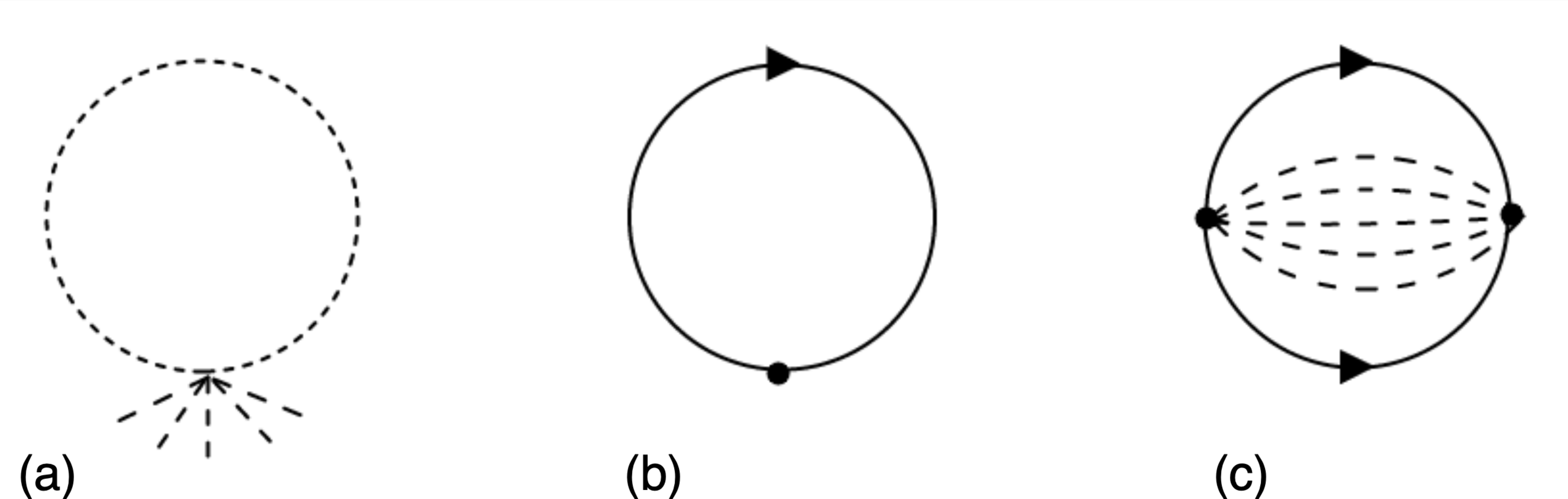}}
\vspace*{8pt}
\caption{List of divergent diagrams: (a) scalar tadpole; (b) fermionic tadpole ; (c)
two-vertex fermionic loop, dressed by an arbitrary number of internal scalar propagators.
\protect\label{fig1}}
\end{figure}

The degree of divergence of a loop diagram 
$D_\Lambda$ is easily computed: \cite{zappalast}
\begin{equation}
D_\Lambda =6\, \left (1- \sum_n   V_n \right)
\, ,
\label{dod}
\end{equation}
where $V_n$ is the number of vertices with $n$ legs (proportional to $g_n$). We notice that $D_\Lambda$
does not depend on the number of external legs and it is always negative unless the diagram has 
one vertex only,  $V_{\widehat n}=1$ and  $V_{n}=0$ for $n\neq \widehat n$, regardless the specific value of $\widehat n$.
Then, one concludes that the only primitively divergent scalar diagrams are those with one vertex, an arbitrary 
number of external legs, and an arbitrary number of tadpoles, all insisting on the same vertex. 
As an example,  the diagram with one tadpole and five external legs  is displayed  in Fig. \ref{fig1} and 
labelled with  (a).

The first consequence of this analysis is that loop diagrams with two or more vertices are not divergent and, since only diagrams with more than 
one vertex can carry dependence on some external momentum, it follows that  quantum corrections depending on external momenta must be finite.
In particular, corrections to the parameters $a_k$ in Eq.~(\ref{ssk}) come from external momentum dependent two-point diagrams and, therefore,
they are finite. An explicit calculation in \cite{zappalast}, shows that these corrections, at least at energy scales above $M$,
are certainly negligible.

The second  consequence comes from the observation that also all operators proportional to  $w_{m,k}$ get quantum corrections from diagrams 
that depend on some external momentum and therefore, when the constraint  $w_{m,k}=0$ is enforced in~(\ref{ssi}), no significant divergent 
correction to this constraint is generated. Then, the dangerous corrections to the speed of light are avoided in a consistent way.

The tadpole (a)  in Fig. \ref{fig1}, gives the elementary UV divergence. It is a one loop diagram that, following \cite{zappalast},
can be regularized by adopting the non-Lorentz invariant cutoff on the modulus of the  tri-momentum, which is consistent with the 
non-Lorentz invariant structure of the action in Eq.~(\ref{ssk}). Since we are interested in the UV properties of the theory, we integrate 
the tadpole between two momentum scales $\Lambda_1$ and $\Lambda_2$, with $\Lambda_1>>\Lambda_2>>M$ and we assume 
$M$ to be much larger than any scale appearing in Eqs.~(\ref{ssk}) and~(\ref{ssi})  that, consequently,  will be neglected. 

Then, for the tadpole (a), we get:
\begin{equation}
I_1(\Lambda_1,\Lambda_2,M) = \frac{M^2}{\sqrt{a_3}\, (2\pi)^2}\ln \left ( \frac{\Lambda_1}{\Lambda_2} \right ) 
+ O \left ( \frac{M^4}{\Lambda_2^2},\, \frac{M^2\Lambda_2^2}{\Lambda_1^2}  \right ) \,.
\label{i1}
\end{equation}
where, without any loss of generality, we can put $a_3=1$. \cite{zappalast}

If we now compute the quantum corrections of a generic coupling $g_n$, the combinatorial factors of each contributing diagram always
yield the following series (which corresponds to the exponential series, if the various couplings are neglected):
\begin{equation}
g_n(\Lambda_2) = \sum_{m=0}^{\infty} \frac{g_{_{n+2m}}(\Lambda_1)}{m!} \; \left [ \frac{I_1(\Lambda_1,\Lambda_2,M)}{2 M^2} \right ]^m
\label{series}
\end{equation}
The series in (\ref{series}) is simplified if one assumes that the theory has only one coupling, i.e. $g_n=g$ for all $n$.
In fact, in this case  by using Eq.~(\ref{i1}), one finds
\begin{equation}
g(\Lambda_2) = g(\Lambda_1) \; \exp \left [ \frac{I_1(\Lambda_1,\Lambda_2,M)}{2 M^2} \right ] \simeq
g(\Lambda_1) \; \left ( \frac{\Lambda_1}{\Lambda_2} \right )^{ 1/ \left( 8\pi^2 \right) }
\label{expseries}
\end{equation}
and it is straightforward to check that $g(\Lambda_1) \to 0$  in the limit $\Lambda_1\to \infty$, at fixed $g(\Lambda_2)$,  {\it i.e.}
the coupling $g$ is asymptotically free.

This remarkable result can be easily generalized to the case of different 
$g_n$ if, at  the scale $\Lambda_1$,  the set of all $g_n$ is  bounded by 
$0\leq g_n \leq g$, where $g$ is a real number.
Then, the exponential series in (\ref{expseries}) is greater  than the series  (\ref{series}) term by term, 
and it is straightforward to conclude  $g_n (\Lambda_1) \to 0$  for each $n$, when   $\Lambda_1\to \infty$.
Conversely, if the potential term in Eq.~(\ref{ssk}) contains only a finite set of non-zero couplings $g_n \neq 0$,
the theory is renormalizable,  but no asymptotically free coupling is found. \cite{zappalast}

\section{Fermionic Fields\label{sect3}}

In case of fermionic degrees of freedom, the higher derivative part of the 3+1 dimensional action, with $z=3$, is
\begin{equation}
S_{F}=\!\int\!d^{3}x\,dt \; \bar\psi \left[ i\gamma^0 \partial_0 -\left (b_1 + \, \frac{\partial_j \partial^j}{M^2} \right ) (i\gamma^i \partial_i )-m_f \right ] \, \psi
\label{fsk}
\end{equation}
where the summed indices $j$ and $i$ refer to the 3 space coordinates and,
as for the scalar case, we normalize $S_F$ by setting $b_3=1$ in front of the higher derivative term.
To preserve rotational invariance, the higher derivative part in brackets does 
not contain a $b_2$ term (analogous of the $a_2$ term in (\ref{ssk}) ), although 
the essential  terms, linear and cubic in the space derivatives, are included in  (\ref{fsk}).
The linear term in the time derivative sets the scaling dimension of the fermion field, 
$[\psi]_s=3/2$, which, unlike the scalar field, turns out to be equal to its canonical dimension.

Consequently, if we assume a Yukawa-like interaction with the scalar excitations,
the renormalizable interaction sector of the action is 
\begin{equation}
S_{Y}=  - \!\int\!d^{3}x\,dt \; 
 \sum_{n=1}^{\infty} \, y_n \;  \frac{ \bar\psi \psi \; \phi^n}{n! \; M^{n-1}} \, ,
\label{fsi}
\end{equation}
where all powers of the scalar field are taken into account because 
$[\phi]_s=0$ and,  therefore, all these terms are renormalizable.

The analysis of the degree of divergence of the diagrams generated by $S_Y$ yields
\begin{equation}
D_\Lambda= 6-3\, \sum_n Y_n - \frac{3}{2} E_f
\label{dody}
\end{equation}
where  $Y_n$ is the number of vertices proportional to $y_n$ and $E_f$ is the number of external fermionic legs.
Then $D_\Lambda$ is non-negative if $E_f=2$ and $\sum_n Y_n=1$, or  $E_f=0$ and $\sum_n Y_n=1$, or 
$E_f=0$ and $\sum_n Y_n=2$. In the first case, when $E_f=2$, diagrams have, besides the two external fermionic
legs, all possible multiple scalar tadpoles insisting on the same vertex. 
In the second case, with $E_f=0$, diagrams display  the fermionic tadpole (b) in Fig. \ref{fig1},
 together with all possible multiple scalar tadpoles insisting on the same vertex. 
In the third case,  again with $E_f=0$,   diagrams are of the (c) kind  in Fig. \ref{fig1}, where not only each vertex 
can be dressed with  all possible multiple scalar tadpoles as before, 
but also an arbitrary number of internal scalar propagators can link the two vertices.
In all three cases, an arbitrary number of scalar external legs  coming out of each vertex is allowed. 
Then, in the diagrams (b) and (c) the vertices indicated by thick black dots,
stand for the full sum of scalar tadpoles together with the arbitrary number of scalar external legs.

By adopting the same regularization procedure employed for scalar fields, the fermionic tadpole (b) 
gives the negative result $-4m_f I_1$, and the most divergent part of the diagram  (c), with $r\neq 0$ internal
scalar propagators, is $-2 I_1^{r+1}$ while the pure fermionic one loop diagram (c) without scalar propagators
gives $-4 I_1$. Therefore, in all cases the divergences amount to  powers of the logarithm of the cutoff.

It is now possible to analyze the renormalization of the couplings and we consider the Yukawa sector 
in the simple case in which all  couplings are equal: $y=y_n$.
For our purposes, together with  $a_3=b_3=1$, which is justified by the negligible running of these couplings in the region above $M$, 
we can also take  $a_1=a_2=b_1=0$ without loss of generality. 
The sum of all  diagrams contributing to the renormalization
of  $y$, integrated as before  between the two scales $\Lambda_1$ and $\Lambda_2$, gives
\begin{equation}
y(\Lambda_2) = y(\Lambda_1) \; \exp \left [ \frac{I_1(\Lambda_1,\Lambda_2,M)}{2 M^2} \right ] \equiv  y(\Lambda_1) \; {\cal E}
\label{yren}
\end{equation}
and this leads to the same conclusions discussed after Eq. (\ref{expseries}), i.e. the coupling $y$ is asymptotically free.

The renormalization of the scalar sector is more involved, because of the diagrams in (c). 
In fact, by keeping different $g_n$ for different $n$, and by retaining the leading divergences only,
we find the following  generalization of Eq. (\ref{series}) 
\begin{equation}
g_n(\Lambda_2) = \sum_{m=0}^{\infty} \frac{g_{_{n+2m}}(\Lambda_1)}{m!} \; \left [ \frac{I_1(\Lambda_1,\Lambda_2,M)}{2 M^2} \right ]^m
 - c_n \; y(\Lambda_1)^2 \; {\cal E}^4 I_1
\label{gren}
\end{equation}
where $c_n>0$ is associated with the combinatorial weight of the diagrams in (c) and its 
 value changes with $n$. Then, also the contribution to  $g_n$ changes with $n$. 
 Therefore, if all $g_n$ were equal to a single coupling $g$ as for the case of the Yukawa coupling $y$,
  it would be impossible to renormalize the full structure in Eq. (\ref{gren}).

In order to determine the $\beta$-function of $g_n$, it is sufficient to take the derivative with respect to $\Lambda_2$ in Eq. (\ref{gren}),
by keeping fixed the values of all couplings at $\Lambda_1$, and it is easy to realize that the two terms in the right hand side (rhs) of (\ref{gren})
give respectively a negative and positive contribution, so that the
sign of the $\beta$-function depends on the balance of these contributions and, consequently, on the  particular values of
the couplings, so that it is not uniquely established. 

However,  the following heuristic argument can be put forward: the stability 
of the scalar potential requires $g_n\geq 0$, at least for all $n$ larger than some $\bar n$. Then, for $n\geq \bar n$ the 
first term in the rhs of (\ref{gren}) must be larger than the second. In addition, we noticed that the coupling $y$ is asymptotically free and 
it grows when the momentum scale is lowered and, in case the $g_n$ are not sufficiently large at the scale $\Lambda_1$ to ensure a negative
$\beta$-function, $g_n$  decreases when the momentum scale runs from $\Lambda_1$  towards $\Lambda_2$. Then, if these two scales are sufficiently 
distant, $g_n$ could become negative at the infrared scale $\Lambda_2$, yielding an unstable potential. 
This is avoided if the values of  $g_n(\Lambda_1)$ are sufficiently large to ensure a negative $\beta$-function, at least for all $g_n$ with $n\geq \bar n$.

It must be noticed that, in the analysis of the fermionic sector we deliberately did not include 2n-fermion vertices. Actually, it is easy to show
by means of  the analysis of the degree of divergence of diagrams, that the only vertex of this kind that is renormalizable, is the 4-fermion vertex.
However,  this vertex can always be reduced, through Hubbard-Stratonovich transformation to a Yukawa-like interaction and, therefore, it should not
be treated as a fundamental interaction. 
In any case, the 4-fermion interaction was already studied,  \cite{Dhar:2009dx} and we do not expect 
that its inclusion could substantially modify our conclusions.
 
A comment on the scalar square mass $g_2\; M^2$ ( as defined in (\ref{ssi}) ), whose correction is given in (\ref{gren})
for $n=2$, is in order.  It is evident that, unless a very subtle cancellation between the first and second term in the rhs  of (\ref{gren}) occurs, 
we get a finite non-vanishing correction to the scalar mass from the momentum region between $\Lambda_1$  and $\Lambda_2$, with $\Lambda_1\to \infty$.
The correction is, in any case, proportional to the scale $M^2$. This is analogous to the result obtained for the  scalar case
in \cite{zappalast}, and even if, as expected, the fermionic contribution to the correction occurs here with opposite sign with respect to the scalar, 
an exact cancellation of the two would be quite unnatural.

\section{Gauge Fields \label{sect4}}

The inclusion of  gauge fields can be realized by defining the appropriate covariant derivative, which can be constructed by starting from 
the generalized derivative $\widehat \partial_\mu$ already introduced in Eq. (\ref{fsk}) (here for simplicity, the  coefficients of the various  
derivative terms, analogous of $b_1,\,b_3$,  are taken equal to 1)
\begin{equation}
 \widehat \partial_0 = \partial_0 \; ,  \;\;\;\;\;\;\;\; 
 \widehat \partial_i  = \left (1 + \, \frac{\partial_j \partial^j}{M^2} \right )   \partial_i  \;,
 \label{sqed}
\end{equation}
and by including the gauge field in the usual way : 
$\widehat D_\mu  =\widehat \partial_\mu  -i q A_\mu$. Accordingly, the generalized electromagnetic tensor is 
$\widehat F_{\mu\nu}=\widehat \partial_\mu A_\nu - \widehat \partial_\nu A_\mu$. However, it is straightforward to realize that
$\widehat D_\mu$, which is non-linear in the derivatives,  does not behave properly under gauge transformations, 
so that the corresponding  higher derivative electrodynamics produces gauge violating contributions to the amplitudes, proportional to powers of $p^2/M^2$, 
where $p^2=p^\mu p_\mu$ is some  square momentum  scale, representative of the amplitude. 

A different higher derivative formulation, that is gauge invariant, was  already  proposed  long ago\cite{horava:ym} but, in this case,
 the higher order space derivatives appear in the form $ ( D_h D_j F_{ik}  )^2$ 
(where $D_i$ are the space components of the standard covariant derivative),
which contains terms analogous to those proportional to $w_{m,k}$ in Eq. (\ref{ssi}) for the scalar case. As discussed above,
in order to avoid model dependent modifications of the light cone, we required to reject  these kind of terms and, consequently,
this formulation cannot be accepted.

If we admit that, together with Lorentz symmetry, also  gauge invariance is violated, at least at high energy,
and only recovered at low energies below $M$, we can make use of the definition in (\ref{sqed}) and of the corresponding 
$\widehat D_\mu$, to include the dynamics of the  field $A_\mu$. Then, independently of gauge symmetry, 
the effect of the dynamics of $A_\mu$ in Eq. (\ref{gren}) is, from an effective point of view,
equivalent to the presence of additional scalar degrees of freedom,
 with the consequence of modifying  only some coefficients in (\ref{gren}) but not the overall structure of the 
equation, so that the conclusions  discussed  above are not qualitatively altered.

\section{Conclusions \label{sect5}}

We analyzed a realization of a higher derivative theory in 3+1 dimensions with anisotropy exponent $z=3$ and, in particular, 
we considered the possible generalization to fermionic and vector 
gauge fields of the specific formulation for scalar fields studied in \cite{zappalast}, with the main intent of constructing a
consistent UV completion of these fields and, ultimately, of the Standard Model. 
Therefore, we focused on the high energy region that corresponds to energies much larger than the scale $M$, 
which sets the scale of Lorentz violating effects. 

In fact, as evident from Eqs. (\ref{ssk}) and (\ref{fsk}), all modifications to the standard 
Lorentz invariant dispersion relations of scalars and fermions, correspond to Lorentz violating additive corrections proportional to 
powers of  the ratio  $(p^2/M^2)$, which become relevant only for $p^2>>M^2$. In addition, an acceptable crossover to the
region below $M^2$, where the usual renormalization group flow of the various couplings is recovered  and Lorentz, as well as 
gauge symmetry are effectively restored, has been realized and discussed in \cite{zappalast}.
As already mentioned, astrophysical high energy gamma ray observations put a strong limit on $M$, that,
for models with corrections proportional to $(p^2/M^2)$ as in our case, is about $M\geq 10^{10}$ GeV,\cite{ellis:2003,chenhuang}  a very large value
if compared with the characteristic scales of the Standard Model.

From the point of view of renormalizability, there is a great improvement due to the higher derivative terms.
In particular, the scalar self-interacting sector  has a very simple UV structure that indicates that the couplings $g_n$ are asymptotically free, 
at least when, at the scale $M$, all couplings are positive  and the set of all $g_n$ is bounded  from above; in addition,
if $g_n =g$ for all $n$, the series of significant diagrams, due to their combinatorial weights, sums up to a simple exponential form.

The  interacting system of scalars and fermions reproduces the same structure of divergences in the Yukawa sector with asymptotic
freedom of the corresponding coupling, while for the scalar couplings  the pattern of divergences becomes more complicated.
However, under the hypothesis of stability of the potential, also scalar couplings  turn out to be asymptotically free.
On the other hand, we are not able to include gauge fields without introducing interaction operators that produce physically  unacceptable 
modifications of the light cones for different species of particles. Therefore in this approach, gauge symmetry can only be recovered 
as an emergent low energy  symmetry  (below $M$).

Then, from the phenomenological  point of view, if the violation of both Lorentz and gauge symmetry could be kept under control
by the large value of the lower bound on $M$, the price to pay is that the strong suppression of divergences is limited to the UV region above $M$,
while  standard scaling holds at scales smaller than  $M$. In particular, the correction to the scalar square  mass, although finite, is  proportional
to $g_4 M^2$, that is very large if compared for instance to the Higgs square mass. In this sense, the positive features of this kind of theory, seem to 
have a very weak impact on the physics of the Standard Model at present energies.

\section*{Acknowledgments}

DZ is grateful to A. Bonanno, V. Branchina and R. Percacci for useful comments and suggestions.
This work has been carried out within the INFN project FLAG.

\end{document}